\newcommand{\arxiv}[1]{#1}
\newcommand{\sigcse}[1]{}

\newcommand\full[1]{}

\arxiv{
\documentclass[11pt]{article}
\usepackage{fullpage}
}

\sigcse{} %

\usepackage{hyperref}
\usepackage{graphicx}

\usepackage{xspace}

\usepackage{alltt}
\usepackage{xcolor}
\definecolor{codegreen}{rgb}{0,0.6,0}
\definecolor{codegray}{rgb}{0.5,0.5,0.5}
\definecolor{codepurple}{rgb}{0.58,0,0.82}
\usepackage{listings}
\makeatletter
\lst@AddToHook{OnEmptyLine}{\addtocounter{lstnumber}{-1}}%
\makeatother
\lstdefinestyle{pystyle}{
    backgroundcolor=\color{white},
    basicstyle=\scriptsize\ttfamily, %
    morecomment = [is]{\#\#\#}{\#\#},      %
    commentstyle=\color{brown},
    keywordstyle=\color{blue}, %
    stringstyle=\color{codepurple},
    numberstyle=\tiny\color{codegray},
    breakatwhitespace=false,
    breaklines=false,
    captionpos=b,
    keepspaces=true,                 
    numbers=left,
    numbersep=5pt,
    numberblanklines=false,
    showspaces=false,                
    showstringspaces=false,
    showtabs=false,                  
    tabsize=2
}
\makeatother
\sigcse{}

\newcommand{\Hex}[1]{\hspace{#1ex}}
\newcommand{\Vex}[1]{\vspace{#1ex}}

\newcommand{\mypar}[1]{\Vex{1.4} \noindent {\bf #1.~}}

\newenvironment{code}{\Vex{0.5}\begin{alltt}\small}{\end{alltt}\Vex{0.5}}
\newcommand\co[1]{\mbox{\tt\small #1}} %
\newcommand\m[1]{\mbox{$#1$}} %

\def\mathify#1{\ifmmode{\mbox{$#1$}}\else\mbox{$#1$}\fi}
\newcommand\NOT{\m{\sim\,}\xspace}
\newcommand\AND{\m{\,\land\,}\xspace}
\newcommand\OR{\m{\,\lor\,}\xspace}

\newcommand\SOME{\m{\,\exists\,}\xspace}
\newcommand\EACH{\m{\,\forall\,}\xspace}

\title{Discrete Math with Programming: A Principled Approach}

\sigcse{}

\begin{document}

\arxiv{
  \author{Yanhong A. Liu \Hex{10} Matthew Castelllana\\
  Computer Science Department, Stony Brook University, Stony Brook, New York\\
  liu@cs.stonybrook.edu, matcastellan@cs.stonybrook.edu
  }
\date{}
\maketitle
}

\begin{abstract}
  Discrete mathematics is the foundation of computer science.  It focuses
  on concepts and reasoning methods that are studied using math notations.
  It has long been argued that discrete math is better taught with
  programming, which takes concepts and computing methods and turns them
  into executable programs.  What has been lacking is a principled approach
  that supports all central concepts of discrete math---especially
  predicate logic---and that directly and precisely connects math notations
  with executable programs.

  This paper introduces such an approach.  It is based on the use of a
  powerful language that extends the Python programming language with
  proper logic quantification (``for all'' and ``exists some''), as well as
  declarative set comprehension (also known as set builder) and aggregation
  (e.g., sum and product).  Math and logical statements can be expressed
  precisely at a high level and be executed directly on a computer,
  encouraging declarative programming together with algorithmic
  programming.  We describe the approach, detailed examples, experience in
  using it, and the lessons learned.

\end{abstract}

\sigcse{}

\sigcse{}

\section{Introduction}

Discrete mathematics is the foundation of computer science.  The central
concepts in it---from logic and reasoning, to sets and functions, to
sequences and recursion, to relations and graphs---are essential mental
tools for modeling real-world objects and developing programming solutions,
whether for basic problem solving or for advanced software development.

At the same time, the core discipline of computer science is program
development.  It is taught in introductory programming sequences, branching
at upper levels to projects in courses such as databases, networking,
security, and especially compilers that have to deal with sophisticated
discrete structures for representing computer programs themselves.

As a result, the two primary courses in computer science 
are typically discrete math and program development, the two 
largest-by-hours 
areas in~\cite[p.37]{2013computercurricula}, but with very
different course activities:
\begin{itemize}
\item The former teaches concepts and reasoning methods, with the help
  of paper and pencil or text formatting tools for writing math
  notations and natural language.
\item The latter emphasizes transforming concepts and computing methods into
  executable programs, overcoming idiosyncratic issues of programming
  languages and systems.
\end{itemize}
Students can see some common concepts underlying both, e.g., sets and
sequences in the former are realized as some collection types in the
latter.  However, there are no direct, precise connections between the two
kinds of activities---the same concepts are studied in completely different
contexts,
and computations involving these concepts are expressed completely
differently in math notations than in most programming languages, especially
Java, the current dominant language used in teaching program development.

Clearly, the instruction of discrete math and programming should be
integrated to connect theory with practice, to let them reinforce each
other, and to help students better understand and master both.  This in
fact has been pursued, in a great deal of prior work and effort, as
discussed in Section~\ref{sec-related}.  What is lacking is a principled
approach for doing this, an approach that covers all central concepts in
discrete math and promotes disciplined uses of these concepts in problem
specifications and programming.

This paper presents such a principled approach. The approach has four main features:
\begin{enumerate}

\item\sigcse{} It covers all central concepts in discrete math, especially including
  the fundamental, challenging topic of predicate logic.

\item\sigcse{} It is based on %
 a powerful language with precise syntax and semantics, 
  directly connecting math notations with programming language notations.

\item\sigcse{} It supports clear and precise specifications of problem
  statements using any combination of the language elements for all
  concepts.

\item\sigcse{} It promotes declarative expression of complex computation
  problems but also supports easy expression of algorithmic steps.

\end{enumerate}
The language we use, referred to as DA in this paper, extends the Python
programming language.  It supports proper logic quantification, as well as
declarative set comprehension (also known as set former and set builder)
and aggregation (e.g., sum and product) over sets and sequences.
Declarative specifications of problem statements are directly executable in
Python with the module for DA extensions, just as algorithmic steps are.

This approach was used in teaching Foundations of Computer Science at Stony
Brook University in Spring 2020, entirely as\sigcse{} extra-credit programming
problems added to regular homework assignments.
Our results and analysis of using this approach support broader deployment,
and we give suggestions for future adoption.

\section{Related work}
\label{sec-related}

Many methods for teaching discrete math have been studied since the early
ages of computer science, e.g.,~\cite{engel1973discrete,power2011teaching},
especially those involving programming, e.g.~\cite{mcmaster2007discrete}.

Naturally, algorithms introduced in discrete math are frequently selected
for implementation in a programming language.
For example, students could be required to implement an algorithm of their
choosing in any programming language~\cite{martin1984role}.
Even an entire follow-up lab course could be developed, e.g., for students
who are already familiar with languages such as Java~\cite{setzer2009lab}.
However, these programming components do not require writing declarative,
logical specifications that are central to discrete math.

There have been many uses of various declarative languages, not only
functional and logic languages, but also the SQL database language.
Examples include using FP, ML, and Prolog in a complementary
fashion~\cite{hein1993declarative},
using SML extensively to write recursive functions over lists for many set
and logic operations~\cite{vandrunen2017functional},
and using SQL to program with sets and relations and especially its
\co{EXISTS} operator~\cite{remshagen2010making}.
All of these efforts had to get around the lack of real universal and
existential quantifiers. 

Other approaches used dedicated logic and modeling systems.
For example, an automated system could take a list of facts, and generate
a list of support facts to give students insight into how
first-order logic works~\cite{nohl2007using}.
The Alloy modeling language~\cite{alloy} is excellent
for writing specifications using sets, relations, and predicate logic and
then finding models that satisfy a specification~\cite{ureel2016discrete},
although it suffers from issues with recursion.
The powerful SMT solver Z3 was also used, for solving puzzles~\cite{hong2020using}.
These are farther away from introductory programming.

There are also many studies of using supporting tools,
especially visual tools and more powerful proof tools.
For example, specialized programs were used for visualizing graphs and algorithms that do
depth-first search, etc.~\cite{berry1997improving},
for learning rules in solving visual logic
puzzles~\cite{cigas2005teaching},
for proof editing with helpful checks~\cite{bjornsson2017proof}, 
and for giving meta-level support~\cite{maniktala2020deep}.
There were also efforts that encourage students to make their own tools, e.g., a proof
checker of a natural deduction system, a database management system, a
propositional logic proof system, and a symbolic execution
engine~\cite{li2019towards}.

Additionally, dedicated books have been written on the topic of teaching
discrete math with a programming language, e.g., C~\cite{ahoullman994fcs}
and Python~\cite{stavely_2014,romeo_2018}.
However, these do not cover writing logical and declarative specifications
for the central topic of predicate logic, instead opting for writing
iterations or recursions for traditional written exercises on the subject, or
avoiding the subject altogether.

\section{Approach}
\label{sec-approach}

We first discuss all central topics that a principled approach must cover.
We then describe the use of a powerful and precise language, the support
for clear and precise problem specifications, the fostering of declarative
as well as algorithmic programming, and the use of programming as an
enhancement.

\mypar{Central topics in discrete math}
Discrete math is typically one of the first two courses in computer
science, the other being introductory programming.  Despite many
textbooks written and used for the subject, the central topics are
well-known, as captured in commonly used textbooks,
e.g.,~\cite{epp2011app,epp2011intro,rosen2012dm,chartrand2011dm}, as well
as~\cite{2013computercurricula}.
\begin{itemize}

\item Logic.  This includes propositional logic\full{ (logical operators
    and truth values, validity of arguments and inference rules)}, the more
  general predicate logic that includes quantifiers, and proof methods.

\item Sets, functions, and relations.  This includes definitions,
  operations, and properties
  over sets, functions, and relations.

\item Sequences and recursion.  %
  This includes definitions of sequences, %
  summation and product forms, recursive formulas, and proofs by induction.
\end{itemize}
Additional topics are often included, but they are generally special,
expanded cases of the core topics.  Typical such topics are:
\begin{itemize}

\item Graphs and trees.  A graph is essentially a set of vertices plus a
  binary relation on the vertices.  A tree is simply a graph where each
  vertex has one incoming edge and multiple outgoing edges forming no
  cycles.

\item Counting and probability.  Counting corresponds to the cardinality of
  sets of interest.
  Probability is essentially the count of elements of interest divided by
  the count of all elements.

\end{itemize}
All topics above include aspects of reasoning and proofs, as well as
computations and algorithms.  These aspects are often included in the
expositions of the topics above, e.g., proofs with logic
in~\cite{rosen2012dm} and both proofs and algorithms with many topics
in~\cite{epp2011app,epp2011intro}.  These aspects are sometimes also
covered as separate topics, e.g., proofs
in~\cite{2013computercurricula,chartrand2011dm} and algorithms
in~\cite{rosen2012dm}.
In the expositions of all these topics, examples and applications from
number theory are often used~\cite{rosen2012dm,epp2011intro}, and
corresponding concepts such as Boolean algebra are often
introduced~\cite{rosen2012dm,epp2011intro,chartrand2011dm}.

Among all topics in discrete math, logic is typically viewed as the most
fundamental topic---it is usually the first topic to
study~\cite{epp2011intro,rosen2012dm,chartrand2011dm} and is also
emphasized particularly as driving the entire
subject~\cite{gries2013logical}.
Therefore, a principled approach for discrete math with programming must
cover logic as well as all other topics.

\mypar{Powerful and precise language}
To cover all central topics, one can see the variety of discrete
values, operations on them, and properties about them that must be
expressed.
To express all of them precisely and unambiguously, and to connect them with
programming, a powerful language with precise syntax (i.e., forms) and
semantics (i.e., meanings) is needed.

We describe the use of such a language that combines the advantages of the
two kinds of languages used in introductory courses:
\begin{itemize}

\item Traditional mathematical notations used in discrete math.  These
  notations are
  high-level and concise.  However, they generally do not have formal
  semantics, and allow loose usage
  with no automated checking for syntax or semantics.

\item Programming languages, such as the dominant language Java, used in
  introductory programming.  These languages have precise syntax and
  semantics and are automatically checked for the syntax and executed
  following the semantics.
  However, they are lower-level, tedious, and verbose.

\end{itemize}
The language we use, DA, extends Python. Python is well-known for being
significantly more concise and higher-level than languages like Java and
C/C++, and is already widely used by scientists and high-schoolers alike
and taught to non-CS and novice students, demonstrating its power and ease
of use.

Python already supports sets and sequences, comprehensions over them, and
generator expressions with operators \co{all}, \co{any}, \co{sum},
\co{max}, etc.
on top of commonly used loops and recursive functions for programming at a
high level like pseudocode.
The DA extensions we use support the following main language constructs
that are not in Python, but are essential for expressing all central
concepts clearly and directly.
\begin{itemize}

\item Proper universal and existential quantifications.  These capture the
  exact meaning of quantified statements (that use universal quantifier
  \m{\forall} and existential quantifier \m{\exists}) in predicate logic.

\item Comprehensions over sets and sequences with logic/pattern variables.
  These correspond to set builder notations for forming expressions over
  sets, relations, and sequences.

\item Aggregations over sets and sequences with logic/pattern variables.
  These are similar to comprehensions but support summation, product,
  counting, maximum, and minimum.

\end{itemize}

The quantification forms are built on the best previous languages that
support quantified expressions,
SETL~\cite{kennedy1975intro,Schwartz:Dewar:Dubinsky:Schonberg:86}, designed
exactly as a set-theoretic programming language, and
ABC~\cite{geurts1990abc}, designed exactly to teach introductory
programming.\footnote{In fact, SETL is one of the earliest and most
  powerful programming languages.
  ABC is a descendant of SETL, and Python is a descendant of ABC and
  C~\cite{vanRossum1993introduction}.}  These are discussed in detail in
Section~\ref{sec-logic}.

Use of what we call logic variables, or pattern variables, in comprehension
and aggregation, as well as quantification, was motivated by
a history of informal use in writing declarative set
expressions, e.g.,~\cite{LiuTei95Inc-SCP,Liu+06ImplCRBAC-PEPM,Liu13book},
which led
to its 
precise formalization with patterns by Liu et
al~\cite{Liu+12DistPL-OOPSLA,Liu+17DistPL-TOPLAS}.
These are discussed in Section~\ref{sec-others}.

Together, the extended language DA supports simpler and clearer
problem specification as well as expression of computations and algorithms.
In particular, logical statements 
about mathematical concepts can be directly executed for computation and
checking, unlike math notations on paper that are disjoint from low-level
programs executed on computers.

The only aspect not supported in 
DA is formal development of complete proofs, but such proof development is
well-known to be challenging even for the best experts.  Support for easier
writing of proofs remains a direction for future
study~\cite{schwartz2011comp}.

\mypar{Clear and precise problem specification}
With a powerful language for expressing logical statements, problem
specifications can be written more easily and clearly.
Whether for computations or for proofs, precise problem specification is
the most critical task.  The language we use supports such specification
for any aspect that needs it.
\begin{itemize}
\item Input specification.  This specifies all sets, functions, etc.\
  given, plus logical statements specifying additional relationships among
  the given structures.

\item Output specification.  This similarly specifies the sets, numbers,
  etc.\ to be produced, plus logical statements specifying how the output
  is related to the input.

\item %
  Auxiliary value specification.  This specifies auxiliary sets, functions,
  etc.\ to use, plus logical statements relating them to input, output, and to
  each other.
\end{itemize}
The language constructs we use for quantification, comprehension, and
aggregation can be arbitrarily combined in writing logical statements.
Section~\ref{sec-logic}
gives example specifications.

\mypar{Declarative as well as algorithmic programming}
Dominant programming languages such as Java and C/C++ do not support
quantification, comprehension, or aggregation, so they must be programmed
using iteration or recursion.
For commonplace programming taught in courses and used in practice, the
most important language constructs for expressing computations are
iterations carried out in loops, where assignments and conditions are used
to set and test values for starting and ending the loop and to update
variables in between.

Languages like Python with DA extensions support quantification,
comprehension, and aggregation as built-ins, which are compiled into loops
automatically for execution.
When a specification specifies an output using these constructs, which can
be executed automatically, the specification serves as a good example of
higher-level, declarative programming.
Section~\ref{sec-logic}
gives examples.

\mypar{Programming as enhancement}
When teaching discrete math with programming in the DA language, all
discrete structures, operations on them, and properties about them can be
expressed directly and precisely, and then executed directly.  The
programming part is exactly to write these precisely and directly.

To ensure that students still learn and build their full skills at least as
well as they learn discrete math without programming, we used three general
methods: (1) assign the programming part several days later than the
non-programming part, (2) give in-class exercises for doing the most
important of these problems before the programming part was assigned,
and (3) assign more or larger problems for the programming part so as to benefit more from automatic execution.

Method (1) allows students to do homework problems first in their head.
Method (2) forces students to do, or at least try, homework problems
earlier.  Method (3) helps show that programming is an enhancement.

\section{Predicate logic with programming}
\label{sec-logic}

The topic on logic starts with propositional logic.  It is about using
propositions\full{ (i.e, names representing simple logical statements whose
internals are ignored)} and logical operators\full{ (i.e., forming more complex
statements using conjunctions, disjunction, negation, and implication)}.
Programming with these is straightforward\full{ and is subsumed by programming
with predicate logic}.
\begin{itemize}

\item Propositional logical statements can be directly and precisely
  written in commonly-used programming languages: use program variables for
  propositions, and use Boolean operators for conjunction, etc.

\item A small issue is that implication (e.g., \co{p implies q}) is not
  in common programming languages, but it can be easily defined or directly
  written using its equivalent form with negation and disjunction (i.e.,
  \co{not p or q}).

\end{itemize}
Predicate logic extends propositional logic to include the use of
predicates with arguments and quantifiers quantifying over values of these
arguments.  It is significantly more sophisticated and cannot be expressed
directly in most programming languages, 
not even Prolog.  Our approach for teaching predicate logic with
programming is as follows:
(1) introduce the language, emphasizing quantifications,
(2) write specifications, especially through examples, and
(3) execute the specifications, directly in Python with DA extensions.

\mypar{Language}
Logical operators in Python are used, because they are simple and easy to
read: \co{and} for conjunction (\AND in math notation), \co{or} for
disjunction (\OR), and \co{not} for negation (\NOT or \m{\neg}).

\begin{figure}[t]
  \centering
  \Vex{-1.2}
  \rule{\sigcse{}\textwidth}{.1mm}\Vex{.7}
  \includegraphics[width=\sigcse{}\arxiv{0.5}\textwidth, trim={1.3cm 0.9cm .3cm 1.7cm}, clip]{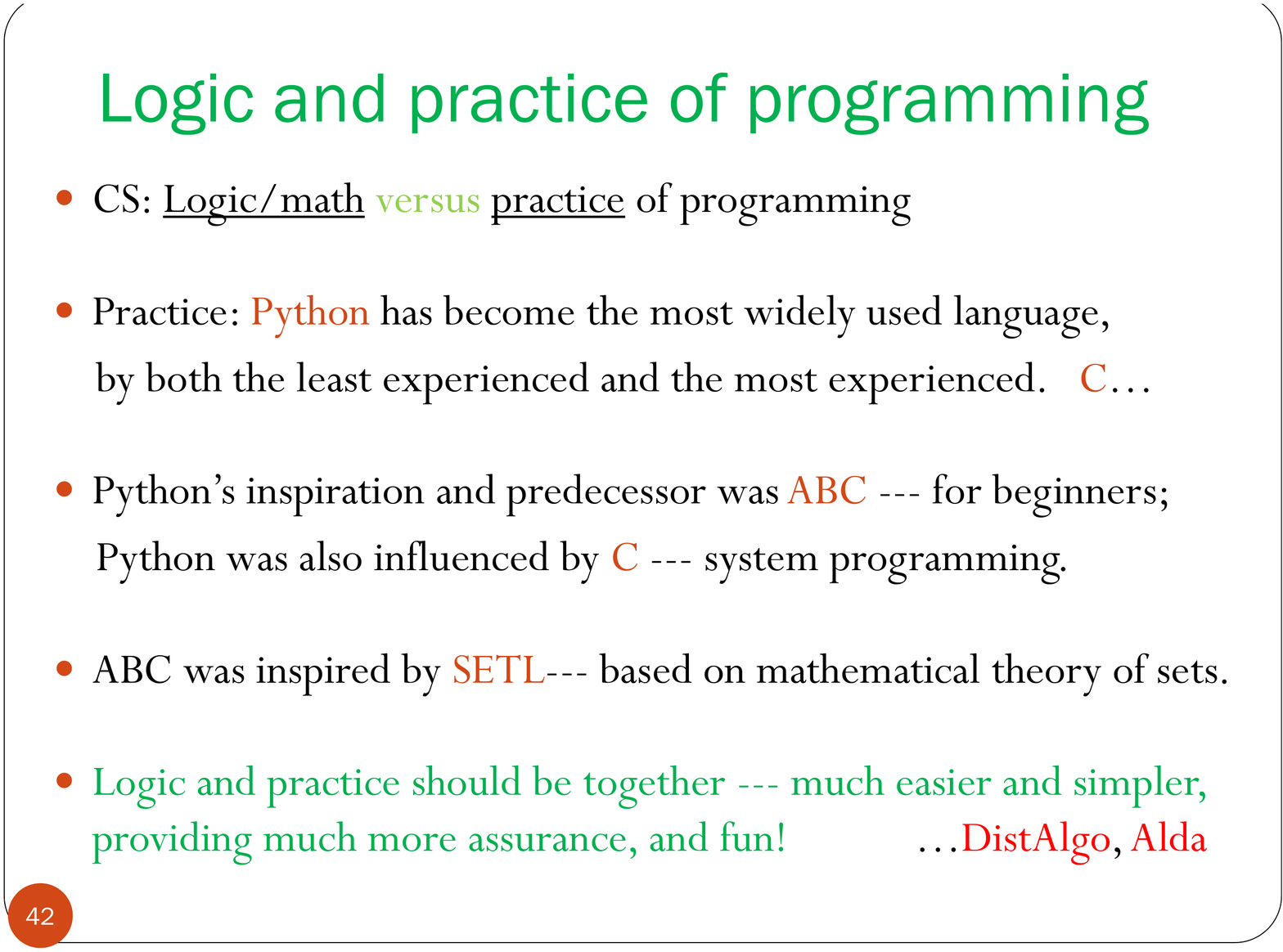}%

 \includegraphics[width=\sigcse{}\arxiv{0.5}\textwidth, trim={1.3cm 4cm .15cm 1.55cm}, clip]{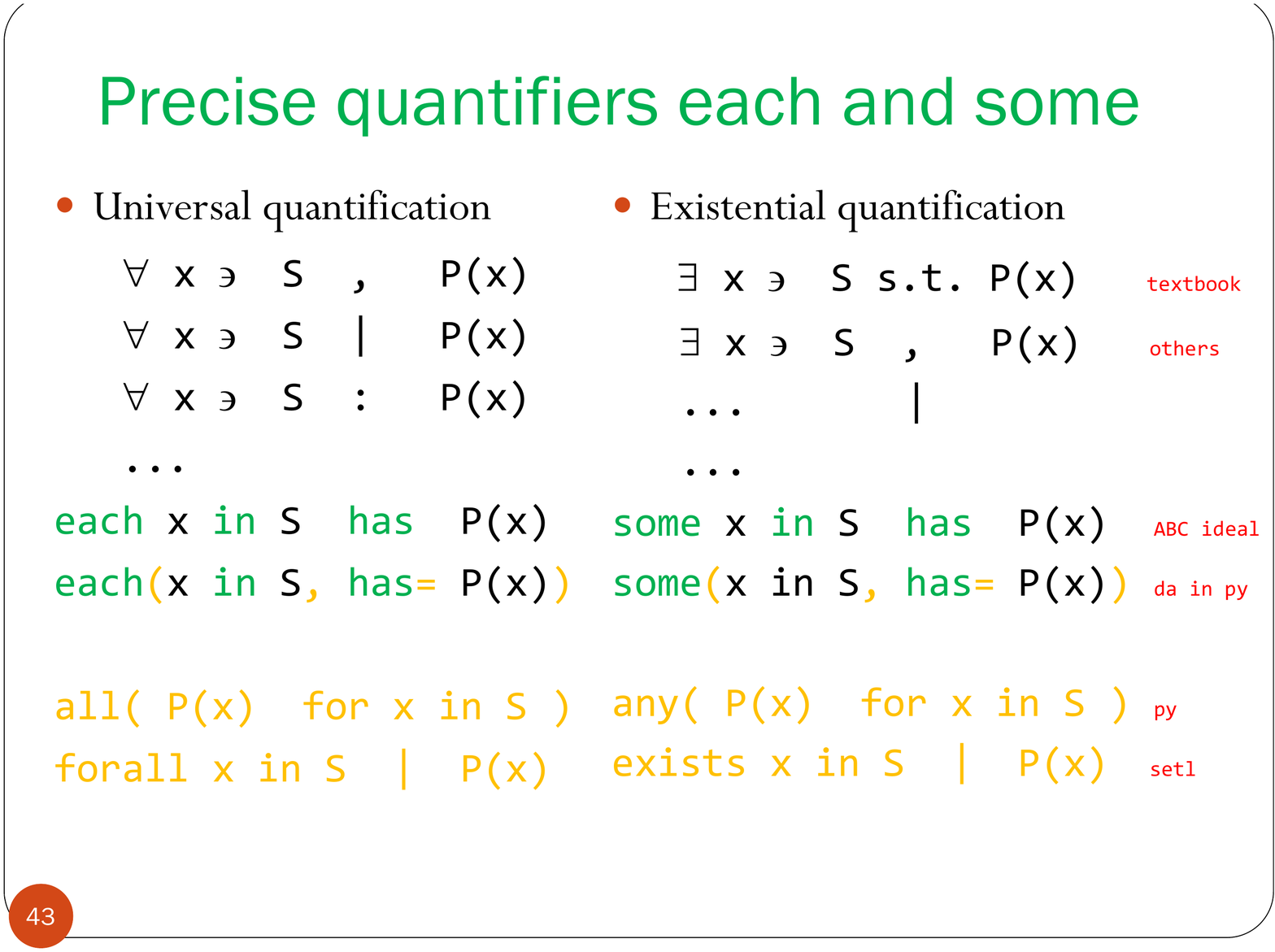}  
  \Vex{-2}
  \rule{\sigcse{}\textwidth}{.1mm}
  \caption{Slides introducing precise quantifications and relationships.\Vex{-2}}
  \label{fig-quant}
\end{figure}

Quantifications, for writing statements with universal and existential
quantifiers (\EACH and \SOME, respectively), are first discussed in class
as usual. Then, only two additional slides, shown in
Figure~\ref{fig-quant}, are discussed.

The first slide gives an overview relating logic and practice of
programming, including programming languages related to Python.  In
particular, Python and C are widely-used languages in practice, and
Python's roots ABC and SETL were meant for beginners and actually based on
set theory, respectively.  It ends with the name of the language to be
used, DistAlgo~\cite{Liu+12DistPL-OOPSLA,Liu+17DistPL-TOPLAS}, plus a
tentative
new name, Alda, for it, abbreviated for both as DA in this paper.

The second slide shows the precise language constructs for quantifications:
the universal quantification means that each element \co{x} in set \co{S}
has property \co{P(x)}, and the existential quantification means that some
element \co{x} in \co{S} has property \co{P(x)}.
\begin{itemize}
\item The first line with \EACH and \SOME is from the
  textbook~\cite{epp2011intro,epp2011app} used for the course.  The second
  and third lines show a few other math notations.

\item The first line with \co{each} and \co{some} is from ABC and ideal for
  reading.  The next line is the form in DA as implemented in Python.

\item The last two lines show the forms in Python and SETL\full{, respectively}.
\end{itemize}
All ABC, DA, and SETL forms match the math notations better than the Python
form.  More critically, the constructs in ABC, DA, SETL, and informally in
math notations---but not in Python---also give a witness:
\begin{quote}
  When the existential quantification is true, variable \co{x} is bound to a
  value in set \co{S} that makes \co{P(x)} true.
\end{quote}
This powerful feature is important for expressing search using math and
logic at a high
level~\cite{kennedy1975intro,Schwartz:Dewar:Dubinsky:Schonberg:86}.

We see that the second slide directly and precisely connects the many
different math notations with the best, easy-to-read programming language
constructs.

No more study of Python or DA was done in class, for three reasons.  (1)
Time diverted from teaching all regularly taught materials should be
minimized.  (2) The homework gave program files that contained examples.
(3) We were confident in the power and ease of Python and DA extensions from
past teaching experience.

\mypar{Specification}
We show the use of DA quantifications in specifying examples with different
combinations.  Two main examples are used in the textbook: a college
cafeteria with students choosing items at different stations, and Tarski's
world as a grid of blocks of various colors and shapes.  We name them \co{cafe}
and \co{tarski}, respectively, and use parts of \co{cafe} as examples.

For \co{cafe}, the textbook provides a figure with example students, food stations,
items in those stations, and the items each student chose.  It then lists four
statements in math notations and discusses their truth values in English. The first example has:
\begin{center}
  \SOME an item \m{I} such that \EACH students \m{S}, \m{S} chose \m{I}.\\
  ``There is an item that was chosen by every student.''
\end{center}
We write the corresponding precise statement as an example in the program
file given to students:
\begin{code}
  some(I in items, has= each(S in students, has= chose(S,I)))
\end{code}

The homework then asks that several statements written in math notations or English
be written in DA.  For example, an exercise problem in the textbook was
used in the homework, asking for the truth values of a list of statements. The first statement is:
\begin{center}
  \EACH students \m{S}, \SOME a dessert \m{D} such that \m{S} chose \m{D}
\end{center}
The expected answer is:
\begin{code}
  each(S in students, has= some(D in desserts, has= chose(S,D)))
\end{code}

In total, in the programming part on predicate logic, 5 statements for \co{cafe}
and 2 for \co{tarski} were asked, all involving nested alternating quantifiers
as the example above, and some also involving \co{and}, \co{or}, and
\co{not}.

\mypar{Execution}
The most practically attractive aspect in teaching discrete math with
programming is that programs are executable to give the specified meaning.
We show execution of the statements written, using the \co{cafe} example.

To execute a statement for \co{cafe}, the sets of stations, items, and students with their choices for predicate \co{chose} must
be given.  The given program file for \co{cafe} contains the following
definitions, followed by example quantifications as discussed earlier, plus
printing of the values of the quantifications.
\begin{lstlisting}
# page 87 of textbook: Example 3.3.3.
# given knowledge from the first paragraph and shown in Figure 3.3.2:
salads = {'green salad', 'fruit salad'}
main_courses = {'spaghetti', 'fish'}
desserts = {'pie', 'cake'}
beverages = {'milk', 'soda', 'coffee'}
stations = [salads, main_courses, desserts, beverages]
choices = {
  'Uta': {'green salad', 'spaghetti', 'pie', 'milk'},
  'Tim': {'fruit salad', 'fish', 'pie', 'cake', 'milk', 'coffee'},
  'Yuen': {'spaghetti', 'fish', 'pie', 'soda'} }
students = choices.keys()

# helper set and function, to capture English more easily:
items = setof(item, sta in stations, item in sta)
def chose(student, item): return item in choices[student]
\end{lstlisting}
Later extra-credit programming assignments gradually include writing more
or all of such sets and definitions by students.
Later assignments also tell students to use witnesses, e.g., print the
value of \co{I} in the first precise statement earlier in this section.

The resulting programs as well as the given programs can be executed directly
in Python with the module for DA extensions.  Students are given two files,
\co{cafe.da} and \co{tarski.da}, and two commands to run, \co{python -m pip
  install pyDistAlgo} for installing DA, and \co{python -m da cafe.da} for
running the cafe example.

\mypar{Programming after thinking and preparing}
Programming should be an enhancement to traditional discrete math coursework, 
while minimizing the chores of working
with a computer system.  The homework instructions try to accomplish this.
For the predicate logic part, there are two prerequisites to doing
extra-credit programming.

First, for the \co{cafe} problem, students were given in-class exercises to do
the exercise problem in the textbook that will be used in the programming part,
before the programming part was posted.
This requires them to think about the problem more before programming.  Only 3 of the 5
statements for \co{cafe} were given for the written part of the homework.

Second, students should install Python and be able to run it on a command
line. This was not
needed if students used machines in the computer lab.

\section{Other topics}
\label{sec-others}

The same approach was also used in teaching all other topics.  We highlight
the power of DA extensions to facilitate this.

\begin{figure}[t]
  \centering
  \Vex{-1}
  \rule{\sigcse{}\textwidth}{.1mm}\Vex{.7}
  \includegraphics[width=\sigcse{}\arxiv{0.5}\textwidth, trim={1.34cm 0.5cm 0.5cm 1.7cm}, clip]{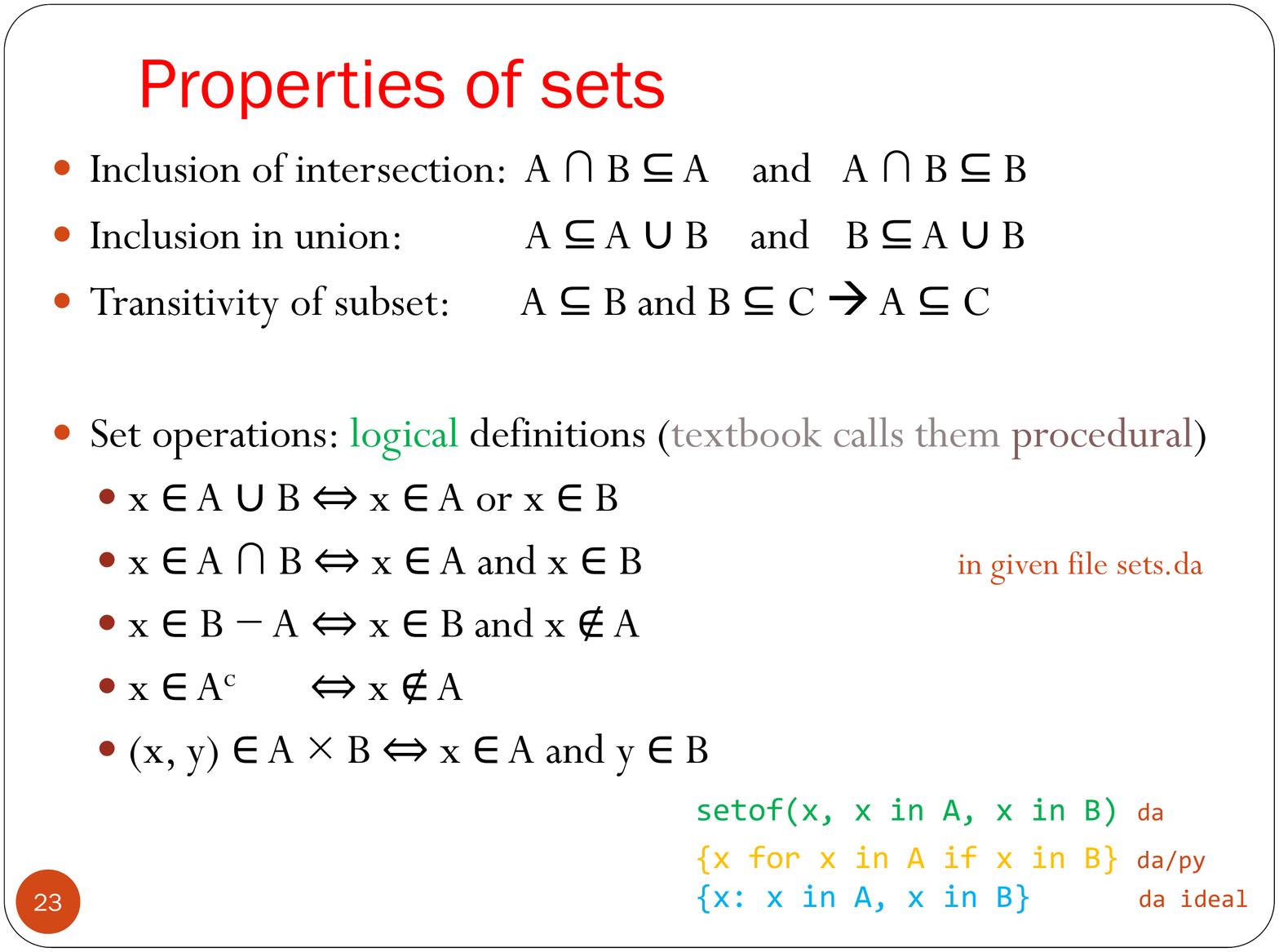}%

  \includegraphics[width=\sigcse{}\arxiv{0.5}\textwidth, trim={1.8cm 3.4cm 0.3cm 1.55cm}, clip]{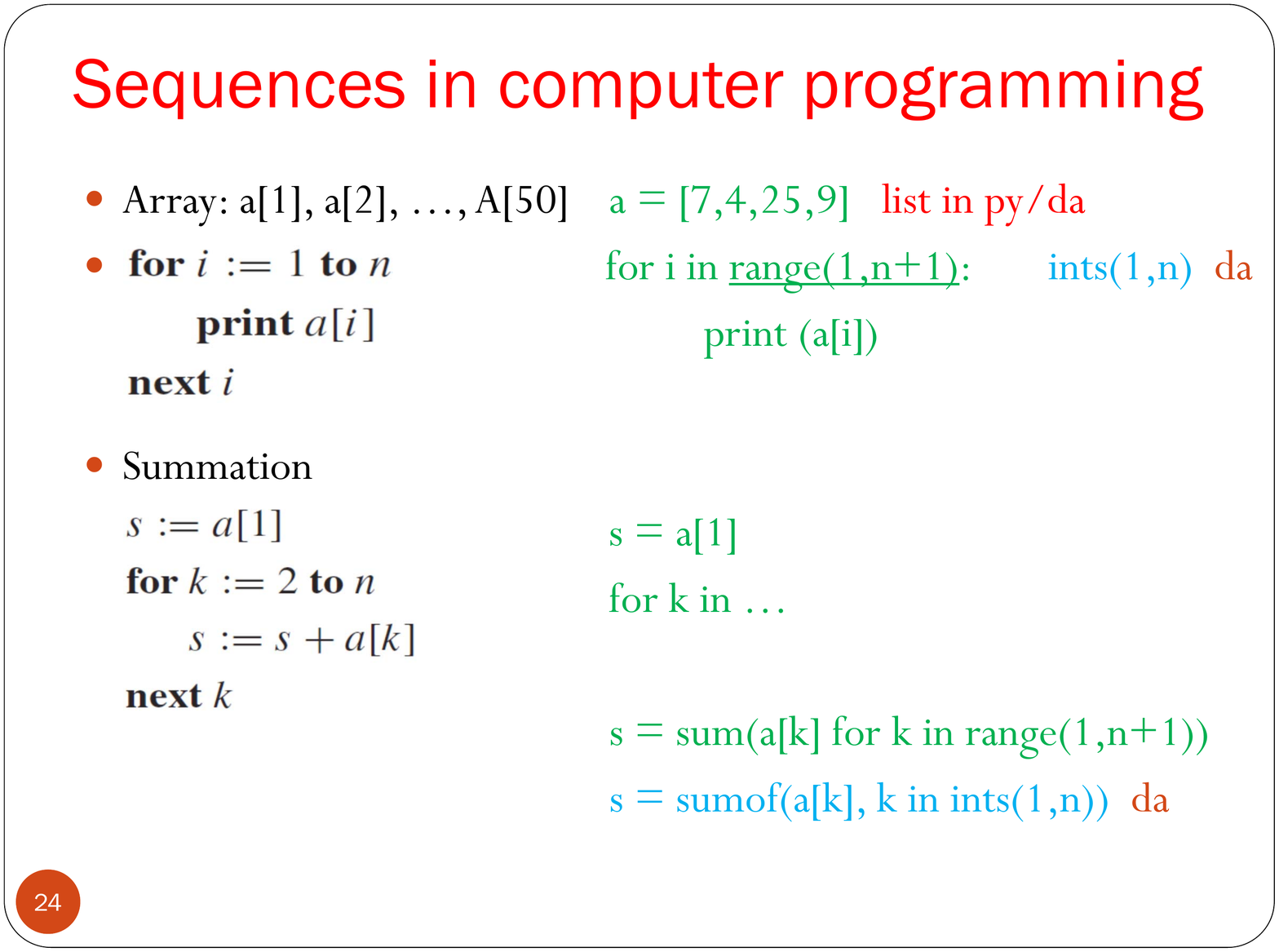}
  \Vex{-2}
  \rule{\sigcse{}\textwidth}{.1mm}
  \caption{Slides on set properties and operations, and on programming with
    sequences.\Vex{-2}}
  \label{fig-others}
\end{figure}

\mypar{Language}
Topics other than predicate logic require simpler language constructs, and
no dedicated slides are used---only one or two slides for each topic with
some notes on the side.  Figure~\ref{fig-others} shows two examples. Most such slides only have
one or two lines of notes on them; these slides are the exceptions.

The first slide shows set operations and properties, all of which can be
directly written in DA.  The notes added on the bottom-right use set
intersection as an example, showing it is in given file \co{sets.da}, and
giving the exact comprehension constructs for expressing it in DA, in
Python (thus in DA too, as DA extends Python), and in an ideal notation.
\begin{itemize}

\item The comprehension in DA means the set of \co{x} satisfying membership
  clauses \co{x in A} and \co{x in B}.  In general, any membership and
  Boolean conditions can be written; the semantics of DA automatically
  avoids the well-known Russell's paradox.

\item In particular, \co{x} is a logic variable, meaning that different
  occurrences of it in a comprehension automatically have the same value.
  So either order of the two membership clauses has the same meaning, and
  the DA compiler can decide how to implement them efficiently.

\end{itemize}
It is critical to note that the comprehension in Python, with \co{for}
followed by \co{if}, is different (and so are comprehensions in SETL): it
means to iterate over elements in \co{A} and, for each element \co{x}, if it
is also in \co{B}, put it in the resulting set.  

Aggregations such as sum and product over sets and sequences can also be
expressed easily in DA.  For example, given a set of sequence \co{S},
\co{sumof(x, x in S)} means the sum of all elements of \co{S}.

The second slide shows pseudocode for programming with sequences.  The
notes added on the right show corresponding constructs in Python, in green,
and alternatives in DA, in blue.  In particular, for computing summation, a
for-loop block can be programmed
as  a 1-line aggregation for sum, in both Python and DA.

In general, Python is both powerful and easy to use and is well-known to be
close to pseudocode when used to program algorithms.  Comprehensions and
aggregations in DA improve over those in Python by being completely
declarative, exactly as in math.

\mypar{Specification}
Using more powerful constructs in DA,
specifications of problem statements and computations for other topics can
also be written easily and precisely as for predicate logic, and be made
executable in Python.

Dozens of extra-credit programming problems were given on expressing
operations on and properties of sets, sequences, functions, and relations,
e.g., writing operations on sequences using both aggregations and recursive
definitions; writing definitions of 1-1 and onto functions and using them
to check given functions; writing definitions of reflexivity, symmetry, and
transitivity and using them to check given relations; writing Euclid
algorithm using both iteration and recursion; writing recursion for Hanoi
Tower; and expressing transitive closure with an existential quantification
with witness in a \co{while} loop.

\mypar{Execution}
All specifications in Python and DA can be executed directly as discussed
in Section \ref{sec-logic}.  The main difference is that, for later topics,
students are asked to write more parts or even all parts of the solution to
a problem on their own.

\mypar{Programming as enhancement}
Besides similar use of programming as for predicate logic, more
computation problems are given, such as Hanoi Tower.  There was even an
extra extra-credit programming part on solving the online exam scheduling
problem that the course itself had, by writing quantifications,
comprehensions, and aggregations in DA and feeding them to a solver.

\section{Results, analysis, and adoption}
\label{sec-results}

To give insight into whether extra-credit programming was truly beneficial, 
we mainly considered two metrics: student test performance, a
quantitative metric; and student surveys, a qualitative metric.
\sigcse{
Additional details can be found in~\cite{LiuCas20DiscreteMathProg-arxiv}.}

\arxiv{
\mypar{Student background and course prerequisites}
	Most students in the course were Computer Science (CS) majors or
	pre-majors.  Most students were in their first year at Stony Brook,
	but most students had some kind of programming experience before
	this course.

	The only prerequisite for the course was basic calculus (Calculus A
	or Calculus I in a series of 3 or 2 courses on calculus).  To do the
	extra-credit programming, students only needed to be able to run
	Python (Python 3.7 was used) on a command line.

	Specifically, 88 out of 115 students total in the class roster were
	listed as CS majors (33) or pre-majors (55).  22 were listed as
	freshman and 75 as sophomore; many in their first year were listed
	as sophomore because they passed a certain number of required
	credits from their first semester and/or their high school
	Advanced Placement courses.

	To better understand the background of the students, a
	questionnaire was given out at the start of the course.  Out of 115
	students enrolled, 108 students responded, including 53 of the 56
	students who did one or more of the programming tasks.

	Out of the 108 respondents, 98 (91\%) had some programming background,
	from a course, job, and/or recreation.
	Of 102 who indicated their class year, 86 (84\%) indicated they were
	in their first year.  Of 106 who indicated their majors, 91 (86\%)
	indicated they were CS majors or pre-majors.

	Out of the 53 respondents who did one of more programming tasks, 49
	(92\%) had some programming background.  Of 51 who indicated their
	class year, 43 (84\%) indicated they were in their first year.  Of
	53 who indicated their majors, 44 (83\%) indicated they were CS
	majors or pre-majors.  These percentages are the same or very close
	to those for the entire class.
}

\begin{figure}[t]
  \centering
  \Vex{-1}
  \includegraphics[width=\sigcse{0.48}\arxiv{.8}\textwidth, trim={0.2cm .52cm .2cm 0.22cm}, clip]{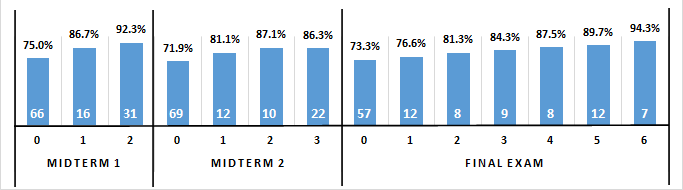}
  \caption{Exam performance vs.\ number of relevant programming assignment
    submissions. The number of students in each category is shown on the
    bottom of the bar.\Vex{-3}}
    \label{fig:ec_vs_exam}
\end{figure}

\mypar{Student test performance}
To see if extra-credit programming helped students learn better, we
examine whether students who did more programming on certain topics
performed better on exams covering those topics.
We consider three groups of programming assignments: the first two
assignments before Midterm 1, the next three before Midterm 2, and all six
before the Final, grouped by their relevancy to the exam.
In each group, students are categorized by how many they submitted for that
group, and the test average from each category was taken.
Figure~\ref{fig:ec_vs_exam} shows the results.

With a very slight exception (0.8\%) in group 2, 
there is a clear positive correlation between the number of programming
assignments submitted and exam performance. Unfortunately, this is not
enough to prove that the assignments by themselves improved student
performance,
due to self-selection bias---it is probable that students who would do
better on exams would do more extra-credit work.
In the future, it may be worthwhile to introduce these assignments in a way
that eliminates self-selection bias.

\mypar{Student surveys}
For each homework, an online survey was created, containing a section
asking for qualitative feedback on extra-credit programming.

For the first programming assignment, on predicate logic,
of 115 students enrolled, 109 submitted the homework;
53 (48.7\%) did not try the extra credit, 20 (18.3\%) tried but had issues
installing or running Python, and 36 (33.0\%) completed all or part of the
programming.
For the 53 who did not try, and the 20 who tried but failed, the optional
nature of the extra credit, being \m{<}1\% of the course grade, mostly likely
did not motivate them enough.
Of the 36 who did programming successfully, 32 (88.9\%) completed both \co{cafe}
and \co{tarski}, and 4 (11.1\%) completed one of them.
Overall, 21 of 36 (58.3\%) indicated they enjoyed the programming, and 24
(66.7\%) indicated they wanted more programming like this.

For comparison, the last programming assignment had 96 students respond;
only 8 (8.4\%) failed to install or run Python---much less than
before---but it also reveals that these students did not try sooner.  44
(46.3\%) did not try the extra credit, and 43 (45.3\%) successfully did one
or more programming problems---again an improvement.  Of the 43 who
succeeded, 32 (74.4\%) indicated they enjoyed the programming---another
positive result.

\arxiv{
\mypar{Follow-up questionnaire}
	A survey was conducted six months after the course ended, with two
	questions: (1) Did you feel that the programming tasks helped you
	learn concepts in the course? (2) Did you feel that the programming
	tasks helped you connect to other CS courses?  The survey was sent
	to the 56 students who did one or more of the programming tasks.
	Out of the 34 respondents, 30 (88.2\%) replied Yes to the first
	question, and 24 (70.6\%) replied Yes to the second.

	It is worth noting that, among the 4 who replied No to the first
	question, 2 were near the top of the class, and 2 were near the bottom.  The large majority of students in the middle, who were more likely to benefit from these activities, felt that the extra credit programming was helpful.

	12 students also wrote additional comments voluntarily.  Most of
	them explained how they liked the programming tasks.  Two mentioned
	that the syntax was confusing at first, but the work was still
	helpful.
}

\mypar{Suggestions for adoption}
The feedback we have received, from both the surveys and in-person
interactions with students, supports that students enjoy discrete math
more with programming.
For students who succeeded installing and running Python, extra-credit
programming was well received.

Clearly not all students come to a discrete math course equally capable of
configuring the minutia of their programming environment.  As programming
was optional, system configuration was only briefly covered in lectures.
For the future, we suggest making programming required to a degree, and
providing in-class configuration sessions.

Once required, the programming part could be given more weight in the
course grade, and could be tested in the exams as well.
Because it was extra-credit programming worth \m{<}1\% of the course grade for
each assignment, and because it came with entirely different and much
longer problem descriptions,
it was harder to initially motivate the average student.
However, this can be easily overcome by increasing its weight in the
course grade. 

Indeed, most students who did not do extra-credit programming indicated they would have liked
to, and some asked about exercises after the course ended.
Already, we have had positive results with 56 students who successfully did
some or all programming problems.  Some went beyond---running distributed
algorithms using DA, asking deep questions about Python, and asking to do
research projects.

We will make problem descriptions and program files for the programming
assignments publicly available.  Programming solutions will be made
available to instructors by requests.

\section{Conclusion}

We have presented 
a principled approach for teaching discrete math with programming by using
a powerful language that extends Python.  The approach and language cover
all central topics,
and allows novice users to understand the concepts precisely, write them
rigorously in specifications, and use them directly in executions.

Our results and analysis of using the approach support broader deployment.
Exploiting
Python also allows the approach to be built on to teach more advanced 
subjects later on, especially with the increasing growth of Python libraries.

\sigcse{}

\arxiv{
\bibliographystyle{alpha}
}
\sigcse{}
\def\bibdir{../../../bib}      %
\arxiv{
\renewcommand{\baselinestretch}{-.1}\small%
}
\bibliography{\bibdir/strings,\bibdir/liu,\bibdir/Lang,\bibdir/Math,refs}

\end{document}